\documentclass[12pt]{article}
\usepackage[utf8]{inputenc}
\usepackage[english]{babel}
\usepackage{amsmath}
\usepackage{mathrsfs}
\usepackage{amssymb}
\usepackage{amsthm}
\usepackage{amsfonts}
\usepackage{xspace}
\usepackage[normalem]{ulem}
\usepackage{graphicx}
\usepackage{multirow}
\usepackage{appendix}
\usepackage{enumerate}
\usepackage{booktabs}
\usepackage{url} 
\usepackage{authblk}
\usepackage{float}
\usepackage{xcolor}
\usepackage{subcaption}
\usepackage{array,makecell}

\usepackage{xcolor}
\usepackage{soul}
\usepackage{tikz}
\usetikzlibrary{arrows,shapes.arrows,shapes.geometric,shapes.multipart,decorations.pathmorphing,positioning,shapes.swigs}
\usetikzlibrary{arrows.meta,decorations.pathreplacing,calligraphy}
\usetikzlibrary{calc}

\usepackage[top=1in,bottom=1.5in,left=1in,right=1in]{geometry}

\newcommand{\bX}{\boldsymbol{X}}
\newcommand{\bx}{\boldsymbol{x}}

\usepackage{multirow}

\usepackage{caption}
\usepackage{subcaption}

\usepackage{float}
\usepackage{rotfloat}
\floatstyle{ruled}
\newfloat{algorithm}{tpb}{loa}
\providecommand{\algorithmname}{Algorithm}
\floatname{algorithm}{\protect\algorithmname}

\usepackage[unicode=true,bookmarks=true,bookmarksnumbered=false,bookmarksopen=false,breaklinks=false,pdfborder={0 0 1},backref=false,colorlinks=true]{hyperref}
\hypersetup{citecolor=blue}
\usepackage{url}

\newcommand\numberthis{\addtocounter{equation}{1}\tag{\theequation}}

\title{Test-negative designs with various reasons for testing: statistical bias and solution}
\author{\small
Mengxin Yu$^{1}$, Tom Hongyi Liu$^{2}$, Kendrick Qijun Li$^3$, Nicholas Jewell$^4$, 

Eric Tchetgen Tchetgen$^1$, Dylan Small$^1$, Xu Shi$^2$, and Bingkai Wang$^2$
\vspace{10pt}

$^1$The Statistics and Data Science Department of the Wharton School, University of Pennsylvania

$^2$Department of Biostatistics, School of Public Health, University of Michigan

$^3$Department of Biostatistics, St. Jude Children's Research Hospital

$^4$Biostatistics Division, School of Public Health, University of California, Berkeley

}
\date{\vspace{-2cm}}
\begin{document}
\def\spacingset#1{\renewcommand{\baselinestretch}%
{#1}\small\normalsize} \spacingset{1}

\maketitle

\begin{abstract}
Test-negative designs are widely used for post-market evaluation of vaccine effectiveness, particularly in cases when randomized trials are not feasible. Differing from classical test-negative designs where only healthcare-seekers with symptoms are included, recent test-negative designs have involved individuals with various reasons for testing, especially in an outbreak setting. While including these data can increase sample size and hence improve precision, concerns have been raised about whether they introduce bias into the current framework of test-negative designs, thereby demanding a formal statistical examination of this modified design. In this article, using statistical derivations, causal graphs, and numerical demonstrations, we show that the standard odds ratio estimator may be biased if various reasons for testing are not accounted for. To eliminate this bias, we identify three categories of reasons for testing, including symptoms, mandatory screening, and case contact tracing, and characterize associated statistical properties and estimands. Based on our characterization, we show how to consistently estimate each estimand via stratification. Furthermore, we describe when these estimands correspond to the same vaccine effectiveness parameter, and, when appropriate, propose a stratified estimator that can incorporate multiple reasons for testing and improve precision. The performance of our proposed method is demonstrated through simulation studies.
\end{abstract}
\noindent%

\spacingset{1.5}
\setcounter{page}{1}
\section*{Introduction}
The test-negative design (TND, \cite{jackson2013test}) refers to a study where patients presenting at healthcare facilities are tested for a disease of interest and categorized as test-positive or test-negative. 
The intervention effect of interest is then estimated as the odds ratio of intervention comparing test-positive and test-negative patients. 
As an observational study design, the TND has been increasingly used to study vaccine effectiveness (VE) for infectious diseases such as influenza \cite{jackson2017influenza}, coronavirus \cite{dean2021covid}, and rotavirus \cite{schwartz2017rotavirus}, especially when randomized trials are infeasible or unethical. Compared to a cohort design, the TND has the advantage of efficiency and cost-effectiveness, and can potentially {reduce} bias arising from differential healthcare-seeking behaviors \cite{sullivan2016theoretical}. 

Classical TNDs only recruit patients who actively seek healthcare and get tested for infection because of disease-related symptoms, but, in many recent applications,  recruitment also enrolls patients with other reasons for testing, particularly apparent in Covid-19 vaccine studies. 
For example, in a TND study evaluating the effectiveness of face masks for preventing Covid-19 \cite{andrejko2022effectiveness}, only 36.4\% patients were tested due to symptoms, accounting for 77.9\% of the positive tests but only 16.7\% of the negative tests. Other reasons for testing included routine screening through work or school (29.9\%), testing required for medical procedures (12.4\%), out of curiosity (12.2\%), and travel screening (8\%). In two other TNDs of Covid-19 that reported reasons for testing \cite{andrejko2022prevention, andrejko2022predictors}, symptoms took a slightly smaller proportion of tested subjects (25.3\% and 29.5\%), while contact with positive cases emerged as a major reason for testing (52\% and 18\%). In addition to these examples, many Covid-19 TND studies involved asymptomatic patients \cite{bruxvoort2021effectiveness}, mandatory testing for healthcare workers \cite{pramod2022effectiveness}, and massive community screening \cite{bernal2021effectiveness}, none of which can be interpreted as testing due to symptoms. Beyond Covid-19, TNDs were also suggested for evaluating VE in an Ebola outbreak \cite{greiner2015addressing, pearson2022potential}, where some participants can be identified and tested due to close contact with known cases. 
Overall, such changes in patient inclusion criteria in TNDs often occur in an outbreak setting, where a great number of tests are not from symptom-related spontaneous healthcare seeking, and these tests are often too frequent to be overlooked.


Despite the emergence of TNDs that include various reasons for testing, {appropriate methods to analyze such studies remain an open question}. The status quo is to ignore the reasons for testing and perform a standard analysis, e.g., via odds ratios and logistic regression, aggregating the data across all reasons for testing. 
However, this method can be biased for estimating the overall VE since different reasons for testing may involve heterogeneous populations, across which the VE, the likelihood of testing, and covariate distributions can all differ.
\cite{lewnard2021theoretical}  pointed out that reasons other than symptoms for testing may introduce bias, and \cite{shi2023current, ortiz2023potential} conceptualized this issue through causal graphs. Both papers and \cite{vandenbroucke2022evolving} stated that a stratified analysis based on testing reasons could alleviate the bias, but no complete quantification or formal solution was given. 
\cite{pearson2022potential} addressed the bias from a naive TND analysis when there were two reasons for testing: (i) symptomatic self-reporting or (ii) testing close contacts of known cases, and proposed a  hybrid design to remove such bias. However, they made a strong assumption that the VE for either recruitment population is the same; in addition, other testing reasons, such as mandatory screening, were not considered in their setting. 
Finally, a naive solution is to filter out asymptomatic subjects and only use the remaining data to perform a classical TND analysis. Although this procedure is valid, it loses a large amount of information.  

In this paper, {we combine directed acyclic graphs (DAGs, \cite{glymour2016causal}) with additional assumptions} to examine the statistical properties of TNDs with various reasons for testing, and propose several estimands with corresponding unbiased estimators. 
First, we revisit the classical TND to clarify underlying assumptions and an appropriate estimand, and {then identify two additional categories of reasons for testing: mandatory screening and contact tracing, for which we construct corresponding DAGs and derive the target estimands. The target estimands for these three categories are generally distinct, which expand the classical scope of TNDs and provide new perspectives on the generalizability of TND estimates. Second, given multiple reasons for testing, we explain why the odds ratio estimator pooling all testing reasons may be biased. Finally, we propose several estimators to target each VE of interest using the corresponding data. Furthermore, we provide conditions to justify a combined estimator based on multiple reasons for testing, thereby maximizing the use of relevant data to achieve unbiased estimation, valid hypothesis testing, and increased precision.}
Moreover, our proposed methods do not solely depend on parametric model assumptions; they also offer the flexibility to incorporate non-parametric kernel smoothing to approximate the underlying data-generating distribution.
To our knowledge, we provide the first vehicle to clarify to what extent asymptomatic reasons for testing can be included in TNDs without introducing bias, and our proposed methods are the first to accommodate a range of reasons for testing, which can inform statistical practice for future TNDs.

\section*{Defintions and assumptions}
We consider TNDs to study a specific disease, namely Covid-19, with a view to estimating vaccine effectiveness. For simplicity, we consider assessing only a single vaccination dose, although the development can be extended to cover situations where $V$ captures a more general vaccination history. Thus, for each subject, 
let $V$ be the vaccination status with $V=1$ if vaccinated and $0$ otherwise. Let $I$ be the true infection status for severe acute respiratory syndrome coronavirus 2 (SARS-CoV-2), where $I=1$ if the subject is infected and $I=0$ otherwise, as determined by a definitive diagnostic test. {Let $S$ be an indicator of having Covid-like symptoms, such as fever and coughing.}  We use $\bX$ to denote {observed} confounders, such as age, sex, and calendar time, and $H$ is a latent (unobserved) healthcare-seeking behavior indicator. Again, for simplicity, we assume $H$ is a binary random variable with $H=1$ indicating {actively seeking care if feeling ill}. {Finally, we define $C$ as a binary indicator of having contact with known cases, which can be impacted by the vaccination status. }

For TND studies involving multiple reasons for testing, we define $R$ as a categorical variable encoding the reasons for testing with mutually exclusive categories, with each category associated with a single reason, such as ``routine screening for work'', or a pool of related reasons, such as ``Covid-like symptoms''. We let $T$ be a binary indicator of being tested at a healthcare setting (and thus eligible for inclusion in a standard TND) with $T=1$ if tested and $0$ otherwise. Throughout, we assume all the above variables are measured without misclassification. 
This setup also applies to studies on other infectious diseases, such as influenza, and other non-vaccine interventions, such as enforcing social distance, with the only difference being that the reasons for testing may be different.

{We assume the joint distribution of the defined random variables is characterized by the DAG in Figure~\ref{fig:complete-DAG}. For the pre-test variables $\mathcal{Z} = \{X,H,V,C,I,S\}$, the assumed causal structure is the same as TND literature \cite{sullivan2016theoretical, shi2023current, ortiz2023potential, schnitzer2022estimands} with the only difference being the new post-vaccination variable $C$, which can directly impact infection and symptoms. We introduce this variable for situations when testing is due to case contact tracing. For testing-related variables $(R,T)$, since $R$ may involve multiple categories of reasons, the joint distribution of $(R,T)$ can depend on all pre-test variables as shown in Figure~\ref{fig:complete-DAG}. However, for a specific reason $r$, $P(R=r,T=1|\mathcal{Z})$ may be relevant to only a subset of $\mathcal{Z}$, and we characterize three categories in the next section.}

    \begin{figure}[htbp]
    \centering
    \begin{tikzpicture}
			\centering
			\tikzset{line width=1pt,inner sep=5pt,
				swig vsplit={gap=3pt, inner line width right=0.4pt},
				ell/.style={draw, inner sep=1.5pt,line width=1pt}}

			\node[shape = circle, ell] (X) at (-4, 2.5) {$X$};
			
			\node[shape = circle, ell] (V) at (-5, 0) {$V$};
			
			\node[shape = circle, ell] (I) at (-1, 0) {$I$};
   
            \node[shape = circle, ell] (S) at (1, 0) {$S$};

            \node[shape = circle, ell] (H) at (-2, 2.5) {$H$};  

            \node[shape = circle, ell] (C) at (-3, -1) {$C$};  

            \node[shape = circle,align=center,ell] (R) at (3, 0) {$R$};

			\node[shape = circle, ell] (T) at (4.5, 0) {$T$}; 
			
   \draw[-Latex, line width=0.5pt, bend right, out= -60, in=-120](V) to (S);

			
			\foreach \from/\to in {X/V, X/I, X/S, V/I, R/T, I/S, X/H,H/V,H/I, V/C,C/I,C/S, X/C,H/C}
			\draw[-Latex, line width = 0.5pt] (\from) -- (\to);

            \draw[draw=blue] (-5.5,-2) rectangle (1.5,3);
            \node[text=blue] (Z) at (-2, -2.25) {Pre-test variables $\mathcal{Z}=\{X,H,V,C,I,S\}$}; 

            \draw[-Latex, line width=0.5pt,draw=blue](1.5,0) to (R);
            \draw[-Latex, line width=0.5pt,draw=blue,bend right](1.5,0) to (T);
            
		\end{tikzpicture}
\caption{The data generating process described by a DAG. The blue arrow is a simplified presentation of arrows from each node in $\mathcal{Z}$, e.g., $\mathcal{Z} \textcolor{blue}{\rightarrow} R$ stands for six arrows from all pre-test variables to $R$. Here, $X,H,V,C,I,S, R,T$ are observed confounders, healthcare-seeking behavior, vaccination, case contact, infection, symptom, reason for testing, and testing, respectively.}\label{fig:complete-DAG}
\end{figure}

{In a TND involving multiple reasons for testing, all individuals tested for SARS-CoV-2 are included in the study. In a prospective design, individuals are recruited, tested, and then categorized as cases or controls. Alternatively, a retrospective design examines electronic health records and extracts data related to testing for analysis. For both implementations, we regard the individual data $(\mathcal{Z}, R)$ with $T=1$ as random samples from the joint distribution characterized by Figure~1. Furthermore, our framework also accommodates a meta-analysis scheme, where individual data are first sampled from $\Pr(\mathcal{Z}|R=r,T=1)$ for each reason $r$ and then data across testing reasons are pooled for analysis.}

\section*{Three categories of reasons for testing}
\subsection*{Testing due to symptoms as in classical TNDs}
{The classical TND only involves individuals who seek care and are tested for infection because of Covid-like symptoms, and we denote them with $R=\textup{Symptoms}$. Therefore, we assume that $$\Pr(R=\textup{Symptoms}, T=1|\mathcal{Z}) = H \times S \times f_s(\bX, V)$$ for some unknown function $f_s$, which means testing due to symptoms implies $H=1$ and $S=1$ (by definition) and can further depend on observed confounders and vaccination status. Here, we assume testing is not driven by the contact history or the true infection status, which implies that this conditional probability is independent of $C$ or $I$. For example, $f_s(\bX,V) = 0.3-0.1V$ means 30\%  chance of testing due to symptoms for unvaccinated symptomatic healthcare seekers, but this probability is reduced by 10\% for vaccinated individuals. With this assumption on $(R=\textup{Symptoms}, T=1)$, the DAG in Figure~\ref{fig:complete-DAG} implies the DAG for testing due to symptoms in Figure~\ref{fig:1a}.}


\begin{figure}[htbp]
    \centering
    \begin{tikzpicture}
			\centering
			\tikzset{line width=1pt,inner sep=5pt,
				swig vsplit={gap=3pt, inner line width right=0.4pt},
				ell/.style={draw, inner sep=1.5pt,line width=1pt}}

			\node[shape = circle, ell] (X) at (-4, 2.5) {$X$};
			
			\node[shape = circle, ell] (V) at (-5, 0) {$V$};
			
			\node[shape = circle, ell] (I) at (-1, 0) {$I$};
   
            \node[shape = rectangle, ell] (S) at (0.7, 0) {$S=1$};

            \node[shape = rectangle, ell] (H) at (-2, 2.5) {$H=1$};  

            \node[shape = circle, ell] (C) at (-3, -1) {$C$};  

            \node[shape = rectangle,align=center,ell] (R) at (4, 0) {$R=\textup{Symptoms}, T=1$};

			
   \draw[-Latex, line width=0.5pt, bend right, out= -50, in=-120](V) to (S);
			\draw[-Latex, line width=0.5pt, bend right, out= -55](V) to (R);

			
			\foreach \from/\to in {X/V, X/I, X/S, V/I, I/S, X/H,H/V,H/I, V/C,C/I,C/S, X/C,H/C, H/R, X/R, S/R}
			\draw[-Latex, line width = 0.5pt] (\from) -- (\to);
		\end{tikzpicture}
        \vspace{-40pt}
\caption{The DAG for classical TNDs. Circle nodes stand for random variables, while rectangle nodes represent that the random variables are conditioned on specific values.}\label{fig:1a}
\end{figure}

Beyond the DAG, we assume that vaccination does not reduce non-SARS-CoV-2 causes of symptoms among healthcare-seekers, i.e.,
\begin{align*}
    {\Pr(I=0, S=1|V=1,\bX=\bx,H=1) = \Pr(I=0, S=1|V=0,\bX=\bx,H=1).}
\end{align*}
{This assumption is commonly used in standard TNDs to reduce an odds ratio parameter to relative risk \cite{jackson2013test, schnitzer2022estimands}.}
These assumptions together motivate an odds-ratio estimator for the VE on the relative risk scale: 
\begin{align*}
    {1- \textup{Odds-Ratio}_{I\sim V}(\bX=\bx, R=\textup{Symptoms}, T=1),}
\end{align*}
where the term $\textup{Odds-Ratio}_{I\sim V}$ stands for the odds ratio of $I$ on $V$ conditioning on the values in the parenthesis.
With mathematical derivation from Web Appendix A, it follows that
\begin{align*}
    &{1- \textup{Odds-Ratio}_{I\sim V}(\bX=\bx, R=\textup{Symptoms}, T=1)} \\
    &={1- \frac{\Pr(I=1,S=1|V=1,\bX=\bx,H=1)}{\Pr(I=1,S=1|V=0,\bX=\bx,H=1)}.} \numberthis\label{VE-rr: symptoms}
\end{align*}
{We denote the right-hand side of Equation~\eqref{VE-rr: symptoms} as {$VE_{symp}(\bx,1)$}, as it evaluates the conditional VE against \textit{symptomatic} infections, i.e. being diseased, among healthcare-seekers. Of note, the conditional VE is a function of $\bx$, which means inference is made on each stratum, such as a specific age group.} 
{Here, classical TNDs achieve VE identification via odds ratio because of its unique design that controls are also healthcare seekers with Covid-like symptoms and that vaccination has no effect on other types of infections, which together cancels out the collider bias from $T$ and target the relative risk.}



\subsection*{Mandatory screening}

{Mandatory screening refers to testing due to massive screening, back-to-school prerequisites, international traveling, or other factors that are irrelevant to their infection status or symptoms.} In other words, tests are motivated by external requirements instead of internal concerns about health. By this definition, there is no direct effect from {$H,C, I,S$} to $R= \textup{Mandatory screening}$ or $T=1$, which implies
$${\Pr(R= \textup{Mandatory screening},T=1|\mathcal{Z}) = f_m(\bX,V)}$$
for some unknown function $f_m$. Here, mandatory screening can be directly impacted by observed covariates and vaccination status since unvaccinated subjects may be subject to stricter testing requirements.

Given this assumption and Figure~\ref{fig:complete-DAG}, we obtain the DAG for mandatory screening in Figure~\ref{fig:screening} and
\begin{align*}
    & 1 - \frac{\Pr(I=1|V=1,\bX=\bx, R = \textrm{Mandatory screening}, T =1)}{\Pr(I=1|V=0,\bX=\bx, R = \textrm{Mandatory screening}, T =1)}\\
    &= 1-\dfrac{\Pr(I=1|V=1, \bX=\bx)}{\Pr(I=1|V=0,\bX=\bx)},
    \numberthis\label{VE-rr: screening}
\end{align*}
{where the left-hand side is an estimable quantity from data under mandatory testing, and the right-hand side defines the conditional VE against \textit{infections} (covering both symptomatic or asymptomatic infections), which we denote as $VE(\bx)$.}

\begin{figure}[htbp]
    \centering
    \begin{tikzpicture}
			\centering
			\tikzset{line width=1pt,inner sep=5pt,
				swig vsplit={gap=3pt, inner line width right=0.4pt},
				ell/.style={draw, inner sep=1.5pt,line width=1pt}}

			\node[shape = circle, ell] (X) at (-4, 2.5) {$X$};
			
			\node[shape = circle, ell] (V) at (-5, 0) {$V$};
			
			\node[shape = circle, ell] (I) at (-1, 0) {$I$};
   
            \node[shape = circle, ell] (S) at (0.5, 0) {$S$};

            \node[shape = circle, ell] (H) at (-2, 2.5) {$H$};  

            \node[shape = circle, ell] (C) at (-3, -1) {$C$};  

            \node[shape = rectangle,align=center,ell] (R) at (4.5, 0) {$R=\textup{Mandotory screening}, T=1$};

			
   \draw[-Latex, line width=0.5pt, bend right, out= -50, in=-120](V) to (S);
			\draw[-Latex, line width=0.5pt, bend right, out= -55](V) to (R);

			
			\foreach \from/\to in {X/V, X/I, X/S, V/I, I/S, X/H,H/V,H/I, V/C,C/I,C/S, X/C,H/C, X/R}
			\draw[-Latex, line width = 0.5pt] (\from) -- (\to);
		\end{tikzpicture}
        \vspace{-40pt}
\caption{The DAG for testing due to mandatory screening. Circle nodes stand for random variables, while rectangle nodes represent that the random variables are conditioned on specific values.}\label{fig:screening}
\end{figure}

{If Covid-like symptoms are also measured in the data from mandatory screening, following a similar procedure, we can also estimate the conditional VE against symptomatic infections by 
\begin{align*}
    & 1 - \frac{\Pr(I=1,S=1|V=1,\bX=\bx, R = \textrm{Mandatory screening}, T =1)}{\Pr(I=1,S=1|V=0,\bX=\bx, R = \textrm{Mandatory screening}, T =1)}\\
    &= 1-\dfrac{\Pr(I=1,S=1|V=1, \bX=\bx)}{\Pr(I=1,S=1|V=0,\bX=\bx)}.
    \numberthis\label{VE-rr: screening2}
\end{align*}
We denote the VE at the right-hand side of Equation~\eqref{VE-rr: screening2} as $VE_{symp}(\bx)$, which differs for $VE_{symp}(\bx, 1)$ in classical TNDs by covering healthcare non-seekers.
}

\subsection*{Case contact tracing}

Case contact tracing is a common practice of monitoring the spread of diseases in an outbreak setting by testing all contacts of known confirmed cases. Because of the mandatory nature of such contact tracing, {testing is unrelated to healthcare-seeking behavior, infection status, or symptoms, but it implies $C=1$ and can be related to covariates and vaccination status. Therefore, we assume $$\Pr(R=\textup{Case contact tracing}, T =1|\mathcal{Z}) = C \times f_c(\bX, V)$$ for some unknown function $f_c$.}
Given these assumptions, we obtain the DAG for case contact tracing in Figure~\ref{fig:case-tracing} and have
\begin{align*}
    & 1 - \frac{\Pr(I=1|V=1,\bX=\bx,  R = \textrm{Case contact tracing}, T =1)}{\Pr(I=1|V=0,\bX=\bx, R = \textrm{Case contact tracing}, T =1)} \\
    &= 1 - \frac{\Pr(I=1|V=1,\bX=\bx, C=1)}{\Pr(I=1|V=0,\bX=\bx, C=1)}, \numberthis\label{eq:ve_cct}
\end{align*}
where the left-hand side refers to an estimable quantity from data and the right-hand side represents the conditional VE against infections among contacts of known cases, which we denote as $VE_{cct}(\bx)$.

\begin{figure}[htbp]
    \centering
    \begin{tikzpicture}
			\centering
			\tikzset{line width=1pt,inner sep=5pt,
				swig vsplit={gap=3pt, inner line width right=0.4pt},
				ell/.style={draw, inner sep=1.5pt,line width=1pt}}

			\node[shape = circle, ell] (X) at (-4, 2.5) {$X$};
			
			\node[shape = circle, ell] (V) at (-5, 0) {$V$};
			
			\node[shape = circle, ell] (I) at (-1, 0) {$I$};
   
            \node[shape = circle, ell] (S) at (0.5, 0) {$S$};

            \node[shape = circle, ell] (H) at (-2, 2.5) {$H$};  

            \node[shape = rectangle, ell] (C) at (-3, -1) {$C=1$};  

            \node[shape = rectangle,align=center,ell] (R) at (4.5, 0) {$R=\textup{Case contact tracing}, T=1$};

			
   \draw[-Latex, line width=0.5pt, bend right, out= -55, in=-120](V) to (S);
			\draw[-Latex, line width=0.5pt, bend right, out= -55](V) to (R);

			
			\foreach \from/\to in {X/V, X/I, X/S, V/I, I/S, X/H,H/V,H/I, V/C,C/I,C/S, X/C,H/C, X/R, C/R}
			\draw[-Latex, line width = 0.5pt] (\from) -- (\to);
		\end{tikzpicture}
        \vspace{-40pt}
\caption{The DAG for testing due to case contact tracing. Circle nodes stand for random variables, while rectangle nodes represent that the random variables are conditioned on specific values.}\label{fig:case-tracing}
\end{figure}

{The above procedure can also be adapted to target the conditional VE against \textit{symptomatic infections} among contacts of known cases by 
\begin{align*}
    & 1 - \frac{\Pr(I=1,S=1|V=1,\bX=\bx, R = \textrm{Case contact tracing}, T =1)}{\Pr(I=1,S=1|V=0,\bX=\bx, R = \textrm{Case contact tracing}, T =1)} \\
    &= 1 - \frac{\Pr(I=1,S=1|V=1,\bX=\bx, C=1)}{\Pr(I=1,S=1|V=0,\bX=\bx, C=1)}, \numberthis\label{eq:ve_cct2}
\end{align*}
which we denote as $VE_{symp,cct}(\bx)$.
}


\subsection*{Summary}

Overall, Table~\ref{tab:summary} summarizes the three reasons for testing, their assumptions, and the target VE discussed above. 
{Under minimum assumptions, each category of reasons for testing targets its own VE estimand. In addition, classical TNDs are constrained to healthcare seekers, whereas ``mandatory screening'' and ``case contact tracing'' cover both healthcare-seekers and non-seekers. Categorization of reasons for testing thus demonstrates the bias from an aggregate odds ratio estimator and also provides a stepping stone for strategically combining reasons for testing, which we introduce below.}

{In the Web Appendix B, we provide the assumptions needed for a causal interpretation of each VE estimand. Among them, it is worth noting that $VE_{cct}(\bx)$ conditions on a post-randomization variable $C$, and hence can be interpreted as the controlled direct effect in causal mediation analysis \cite{richiardi2013mediation}, instead of the standard causal VE as in $VE(\bx)$.}

\begin{table}[htp]
\renewcommand{\arraystretch}{0.8}
    \centering
    \caption{A summary of considered reasons for testing with their assumptions, examples, and target VEs in TNDs. The VE parameters $VE_{symp}(\bx,1)$, $VE(\bx)$,  $VE_{symp}(\bx)$,  $VE_{cct}(\bx)$, $VE_{symp, cct}(\bx)$ and $VE_{cct}(\bx)$ are defined in Equations~\eqref{VE-rr: symptoms}, \eqref{VE-rr: screening}, \eqref{VE-rr: screening2}, \eqref{eq:ve_cct}, and \eqref{eq:ve_cct2}, respectively.
    }
    \label{tab:summary}
    \resizebox{\textwidth}{!}{
    \begin{tabular}{p{2.5cm}p{8cm}p{2.5cm}p{5cm}}
    \hline
    \makecell[l]{Reasons for\\ testing}     &  Main assumptions & Examples &  Target VEs\\
    \hline
    \makecell[l]{Symptoms \\(Figure~\ref{fig:1a})} & \makecell[l]{1. Only symptomatic healthcare-seekers \\ \quad\  are tested.\\
    2. Vaccination does not reduce  \\ \quad\ test-negative symptoms.\\
    3. Given a symptom, the uptake of testing\\ \quad\ is independent of true infection status\\\quad\ or contact with cases.\\}  & Fever, fatigue & \makecell[l]{
   VE against symptomatic \\
   infections for healthcare \\
   seekers: $VE_{symp}(\bx,1)$. }\\
    \hline
    \makecell[l]{Mandatory \\screening\\(Figure~\ref{fig:screening})} &\makecell[l]{1. Testing is not directly impacted \\\quad\ by infection status, symptom, or\\ \quad\ healthcare-seeking behavior.} &\makecell[l]{ Massive tests, \\traveling} & \makecell[l]{
    VE against infection: $VE(\bx)$, \\
    or  VE against symptomatic\\
    infection: $VE_{symp}(\bx)$. }   \\
        \hline
     \makecell[l]{Case contact \\tracing \\(Figure~\ref{fig:case-tracing})} & \makecell[l]{
     1. Only case contacts are tested.\\
     2. Testing is not directly impacted  by\\\quad\ infection status, symptom, or\\ \quad\ healthcare-seeking behavior.} &\makecell[l]{Case contact\\ tracing}  & \makecell[l]{VE against infection among \\case      contacts: $VE(\bx)$, \\
    or  VE against symptomatic\\
    infection among case \\
    contacts: $VE_{symp}(\bx)$.}\\
    \hline
    \end{tabular}
    }
\end{table}


{Beyond the three testing reasons we discussed, $R$ may encode more categories, for which meaningful VE identification may be challenging. For mathematical completeness on $\Pr(R,T|\mathcal{Z})$, we define $R=\textup{Others}$ as the combination of all other testing reasons and denote $\Pr(R=\textup{Others},T=1|\mathcal{Z}) = f_o(\mathcal{Z)}$ for some unknown function $f_o$, which imposes no structure on testing due to other reasons. Finally, we require $f_s, f_m, f_c, f_o$ are non-negative functions with summations uniformly upper-bounded by 1 so that $\Pr(R,T|\mathcal{Z})$ is well defined.} 

\section*{Potential bias given various reasons for testing}



			
			

			

			
			

When a TND involves multiple reasons for testing {(including the above-discussed three categories or other reasons)}, the DAG in Figure~\ref{fig:complete-DAG} can not be simplified. Since the observed data only contain tested individuals,
a naive odds ratio estimator with observed data will be biased for VE, e.g., 
\begin{align*}
    VE(\bx) &\ne 1-\textup{Odds-Ratio}_{I\sim V}(\bX=\bx, T =1)
\end{align*}
since the collider bias arising from conditioning on $T=1$ cannot be removed. This bias also exists for estimating $VE_{symp}(\bx)$ or $VE_{symp}(\bx, 1)$ for similar reasons and, additionally, due to the unmeasured confounder $H$.  As a numerical example, consider $\Pr(H=1)=0.5$, $\Pr(V=1|H)=0.5$, $ \log\{\Pr(I=1|V,H)\}=-0.5-0.5V-1.5H$, and $\Pr(S=1|V,H,I) = 0.5$.
We consider testing due to symptoms, mandatory screening, or other reasons with
\begin{align*}
    \Pr(R=r, T=1|\mathcal{Z}) = \left\{\begin{array}{rl}
    0.2 HS     &  \textrm{if}\ r = \textrm{Symptoms},\\
    0.2V+0.1     &  \textrm{if}\ r = \textrm{Mandatory screening},\\
    0.3VI    &  \textrm{if}\ r = \textrm{Others}.
    \end{array}\right.
\end{align*}
Then, the odds ratio estimator based on aggregated results from tested individuals is $-0.02$, which substantially underestimates the VE among the whole population, which is 0.39. 
{In addition, we have $VE(\bx)=VE_{symp}(\bx)=VE_{symp}(\bx,1) = 0.39$ in this example, which demonstrates that the aggregated odds ratio estimator may not target any VE of interest.}


\section*{{Estimating the VE parameters}}
\subsection*{Unbiased estimation by stratification}
For a TND with multiple reasons for testing, we next show how to estimate the identifiable VEs (summarized in Table~\ref{tab:summary}), based on each type of reason \\$r \in \{\textup{Symptoms, Mandatory screening, Case contact tracing}\}$. For this purpose, we model
$$\Pr(I=1|V, \bX, R=r, T =1) = h(V,\bX,r),$$
where $h(V,\bX,r)$ is a user-specified learning model.
We can assume $h(V,\bX,r)$ follows a parametric model, such as a log-binomial regression model, or a non-parametric model, using kernel smoothing for non-linear effects of $X$; both approaches are well-developed in the literature with software available for estimation. Compared to parametric models, a non-parametric model requires fewer assumptions and can better approximate the truth with the goal of reducing bias, albeit with the disadvantages of requiring larger sample sizes to avoid overfitting and increased variance associated with a slower convergence rate.
We provide mathematical details and model fitting in Web Appendix C. With a user-specified model, we obtain $\widehat{h}(V,\bX,r)$ as the estimate of $h(V,\bX,r)$.

We next construct the following estimators using data from each pre-defined category of reasons for testing: 
\begin{align*}
    \widehat{VE}_{symp}(\bx, 1) &= 1 -\frac{\widehat{h}(V=1,\bX=\bx,r=\textup{Symptoms})\{1-\widehat{h}(V=0,\bX=\bx,r=\textup{Symptoms})\}}{\widehat{h}(V=0,\bX=\bx,r=\textup{Symptoms})\{1-\widehat{h}(V=1,\bX=\bx,r=\textup{Symptoms})\}},\\
    \widehat{VE}(\bx) &= 1- \frac{\widehat{h}(V=1,\bX=\bx,r=\textup{Mandatory screening})}{\widehat{h}(V=0,\bX=\bx,r=\textup{Mandatory screening})}, \\
    \widehat{VE}_{cct}(\bx) &= 1- \frac{\widehat{h}(V=1,\bX=\bx,r=\textup{Case contact tracing})}{\widehat{h}(V=0,\bX=\bx,r=\textup{Case contact tracing})}.
\end{align*}
{The estimators $\widehat{VE}_{symp}(\bx)$ and $\widehat{VE}_{symp, cct}(\bx)$ can be constructed similarly by replacing $I=1$ by $I=1,S=1$ in the definition of $h$, using the data from $r=\textup{Mandatory screening}$ and $r=\textup{Case contact tracing}$, respectively.}

Under our modeling assumptions and regularity conditions listed in Web Appendix C, each estimator converges in probability to their targe VE parameter as the sample size $n$ increases, and their normal-approximated 95\% confidence intervals based on estimated standard error estimators will have correct coverage in large samples. We also provide variance estimators and technical details in Web Appendix C. 

\subsection*{{Connecting VE parameters}}

{In practice, it is common that the above-discussed VE parameters are all distinct. However, with additional assumptions, we are able to connect different VE parameters. For testing due to symptoms, if VE does not vary by the healthcare-seeking behavior \cite{jackson2013test} and $V$ is independent of $H$ given $\bX$, then $VE_{symp}(\bx,1)$ will be equal to $VE_{symp}(\bx)$, which motivates us to combine the VE estimation from different testing reasons. For case contact tracing, we will have $VE_{cct}(\bx) = VE(\bx)$ and $VE_{symp, cct}(\bx) = VE_{symp}(\bx)$, if being a contact of known cases does not modify VE and $V$ is independent of $C$ given $\bX$. Nevertheless, this assumption needs further justification since close contact of cases might represent a higher level of exposure intensity to the virus, a potential confounder for VE.}

{For the VEs we have discussed, it is important to highlight that the VE against symptomatic infection and VE against general infections are different. Since vaccination may further alleviate symptoms beyond guarding against infection, $VE_{symp}(\bx)$ is generally larger than $VE(\bx)$. Therefore, while data from different sources can be combined for a joint analysis,  $VE(\bx)$ and $VE_{symp}(\bx)$ are estimated separately below.}

In fact, whether the above VE parameters are equal is testable from the data. To evaluate whether case contact tracing modifies the VE, we can test the hypothesis $H_0: VE_{cct}(\bx) = VE(\bx)$ for any $\bx$ such that $\widehat{VE}_{cct}(\bx)$ and $\widehat{VE}(\bx)$ can be constructed. Since both $\widehat{VE}_{cct}(\bx)$ and $\widehat{VE}(\bx)$ are normally distributed (on the root-$n$ scale) and independent in large samples, $\widehat{VE}_{cct}(\bx)-\widehat{VE}(\bx)$ also follows a normal distribution in large samples, from which $H_0$ can be rejected if $|\widehat{VE}_{cct}(\bx) - \widehat{VE}(\bx)|$ exceeds the critical value for a given type I error. 
Following a similar logic, we can also test $H_0: VE_{symp}(\bx,1)=VE_{symp}(\bx)$ using $\widehat{VE}_{symp}(\bx,1)$ and $\widehat{VE}_{symp}(\bx)$. This test has greater power if the candidate VEs have substantial differences, the sample sizes are large, and the VE estimates have small variances.


\subsection*{{Combining the stratified estimators to improve precision}}

{In special cases where several VE estimands coincide, we are able to combine the corresponding estimators to yield a more precise estimator. If $VE_{cct}(\bx) = VE(\bx)$, we propose a combined estimator for $VE(\bx)$, defined as}
\begin{align*}
        \widehat{VE}^*(\bx) &= w_1  \widehat{VE}(\bx) + w_2\widehat{VE}_{cct}(\bx),
\end{align*}
where $w_1$ and $w_2$ are arbitrary positive weights such that $w_1 + w_2 = 1$. The weights can be simply the proportion of samples in each data source,  inverse variance-weighting, or other weighting methods from the meta-analysis literature \cite{lee2016comparison}.

{If $VE_{symp}(\bx,1)=VE_{symp}(\bx)=VE_{symp,cct}(\bx)$, we can construct an estimator combining all three categories of reasons:}
\begin{align*}
        \widehat{VE}^*_{symp}(\bx) &= w_1^*\widehat{VE}_{symp}(\bx,1)+ w_2^*\widehat{VE}_{symp}(\bx) + w_3^* \widehat{VE}_{symp,cct}(\bx)
\end{align*}
where $( w_1^*, w_2^*, w_3^*)$ are arbitrary positive weights with $w_1^*+ w_2^*+ w_3^*=1$, and these weights can be chosen similarly to $(w_1,w_2)$. When their assumptions hold, combined estimators improve precision by leveraging the additional data. 
The amount of precision improvement will be greater if we have more data on asymptomatic tests. 
In the Covid-19 context, asymptomatic tests can often account for a large proportion of data, thereby having the potential to improve precision substantially. This result extends \cite{pearson2022potential}, which proposed combining stratified estimators for symptoms and case contact tracing (without specifying how to choose the weights) but did not consider mandatory screening or explain the underlying assumptions for estimation validity. 


\section*{{Numerical illustration}}
\subsection*{Simulation}
{In Web Appendix D, we provide simulation studies to demonstrate our results numerically.  
In summary, the simulation studies show that (1) the traditional aggregated odds ratio estimator can
be biased for TNDs with multiple reasons for testing, (2) the proposed estimators
are unbiased for the target estimands (with the necessary assumptions), (3) leveraging data for asymptomatic reasons for testing can improve precision. Furthermore, supplementary simulations to explore the effect of misclassification of the infection status and missing values in the reasons for testing are also provided in Web Appendix D.}

\subsection*{{Data application}}
{
We use the electronic health records (EHR) within the University of Michigan Health System for a realistic demonstration of our results. Focusing on the period between July 1 and December 31, 2021 (approximately the Delta outbreak), we extract the vaccination history, laboratory results of SARS-CoV-2 tests (positive/negative), Covid-like symptom evaluation (yes/no), and covariates including gender, body mass index, and age. We let $V=1$ if a subject received any vaccine dose. Since reasons for testing are not explicitly recorded in the EHR, we define testing due to symptoms as outpatient testing on symptomatic subjects, and case contact tracing as those with ICD-10 code Z20.822, meaning ``Contact with and (suspected) exposure to Covid-19". For mandatory screening, while the ICD-10 code Z11.52 matches ``screening for Covid-19'', it rarely appeared in the EHR; we therefore exclude this reason for testing from our analysis. It is important to highlight that our definitions of reasons for testing are only for demonstration, which may deviate from the truth and fail in other datasets.}

\begin{figure}[H]
    \centering
    \includegraphics[width=0.8\linewidth]{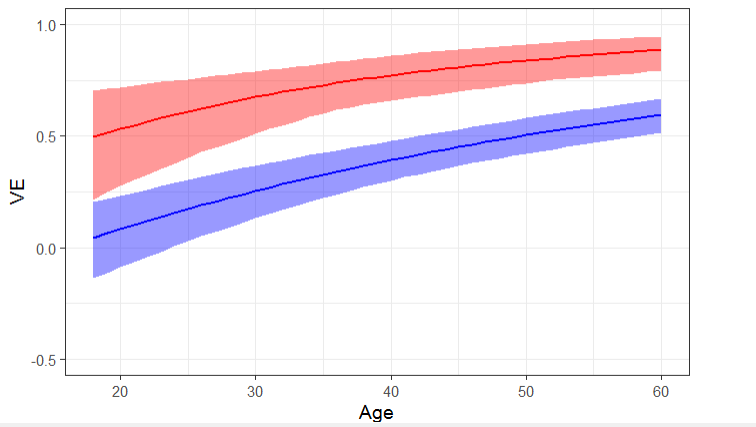}
    \caption{VE by age in the data application. The red and blue colors represent tests due to symptoms and case contact racing, respectively.}
    \label{fig:data application}
\end{figure}

{A parametric generalized linear model is used to estimate $VE_{symp}(\bx,1)$ based on testing due to symptoms and $VE_{symp,cct}(\bx)$ based on case contact tracing. First, we implement the regression model without vaccination-covariate interactions, which results in constant VE estimates across strata. Table~\ref{table: data-application} summarizes the analysis results and shows a significant difference in VEs between the two categories (p-value $<0.001$), which may reflect the effect modification from healthcare-seeking behavior and contact with cases. As a result, we do not perform the combination of the stratified estimators. However, we implemented the odds ratio estimator aggregating both reasons and obtained $40\%$ with standard error $0.04$. This estimator shows bias to either VE, which further demonstrates the issue of an unstratified analysis. Furthermore, Figure~\ref{fig:data application} shows how the VE changes with age through a vaccination-age interaction term in the regression, and the results indicate better VE for older individuals in both categories. }

\begin{table}[H]
\centering
\caption{Summary results of data application.}\label{table: data-application}
\label{tab:delta_firstdose}
\begin{tabular}{rrrrr}
\hline
\textbf{Reason} & \textbf{Sample size} & \textbf{VE} & \textbf{Standard error} \\
\hline
 Symptoms   & 1087  & 0.77   & 0.05 \\
 Case contact tracing   & 22106 & 0.33  & 0.05\\
\hline
\end{tabular}
\end{table}

\section*{Discussion}
As TNDs become increasingly popular for evaluating VE based on observational data, the demand for valid, robust, and appropriate statistical analyses for TNDs is also growing. In this paper, we focus on an emerging issue that a TND may recruit participants with varying reasons for testing, and {clarify the VE estimands associated with the various reasons}. Of course, the suggested approach relies on collecting accurate information on the reason for testing at the time of recruitment participation.
We note that severe bias can be caused if the reasons for testing are ignored. 
In summary, TNDs with various reasons for testing bring not only challenges in analyses but also present opportunities for obtaining precise estimates of VE on different populations. 
Our statistical analysis and proposed methods provide a handle to take advantage of this opportunity and can potentially inform future TND studies.



In designing and implementing TND studies, it is thus important to understand why people are tested. 
If this information is missing, an estimated VE can be biased. In this case, the magnitude and direction of this bias are likely intractable absent further information. Therefore, when tests for asymptomatic subjects are expected, we recommend recording the reasons for testing in patient enrollment or in the surveillance system. For this purpose, an example questionnaire used by the California Department of Public Health in the United States is available in the Supplementary Material of \cite{andrejko2022prevention}.

Our statistical framework assumes that a subject should have a single, accurate reason for testing. In practice, however, a patient may have two or more reasons for testing or, on the contrary, unclear or misclassified reasons for testing. 
While this may not be an issue for mandatory testing such as case contact tracing, bias may occur with self-reporting patients. This issue can be mitigated by adopting survey methods to promote accuracy and honesty \cite{groves2011survey} and can also be examined by a sensitivity analysis. We leave this topic for future research. 

{In order to remove the collider bias on testing, an important assumption we implicitly assume is that the observed covariates include all necessary variables for achieving the conditional independence between $I$ and $(R=r,T=1)$.  Similar to the no-unmeasured confounder assumption in causal inference, evaluating or relaxing this assumption requires additional data or framework. Recently, \cite{kundu2024sensitivity} proposed sensitivity analysis for TNDs, which can be adapted to examine how the results are impacted without this assumption.}

To derive the VE measures given various reasons for testing, we have mainly focused on a simple setting with a binary vaccination status, a binary outcome, {and a binary healthcare-seeking behavior}. In the Covid-19 setting, however, the VE may differ by vaccination doses, the definition of outcomes (infected, diseased, hospitalized, or dead), and the variant of infection. With multi-level $V$ and $I$, our framework can be applied to evaluate the effect of each specific vaccination on each type of outcomes separately, provided that vaccination and outcomes are well recorded and the assumptions in Table 1 hold. {Furthermore, the healthcare-seeking behavior should reflect the patients' propensity to seek care, which is unlikely to be captured by a binary variable. Therefore, the selection bias from testing may be reduced instead of fully removed by conditioning on a binary healthcare-seeking behavior in classical TNDs \cite{lipsitch2016observational,sullivan2016theoretical}.} On this topic, \cite{ciocuanea2021adjustment} proposes measuring symptom severity to alleviate selection bias on the odds ratio scale; non-binary healthcare-seeking behavior is handled as well. In addition, \cite{li2024double} used negative controls to address the bias from healthcare-seeking behavior. A comprehensive examination of these different scenarios is an important future direction.



\subsection*{Acknowledgement}

Research reported in this publication was supported by the National Institute Of Allergy And Infectious Diseases of the National Institutes of Health under Award Numbers R01-AI148127 and R00-AI173395. The content is solely the responsibility of the authors and does not necessarily represent the official views of the National Institutes of Health.

{
\bibliographystyle{ama}
\bibliography{references}
}

\end{document}


\def\spacingset#1{\renewcommand{\baselinestretch}%
{#1}\small\normalsize} \spacingset{1}

\date{\vspace{-5ex}}

\maketitle

\spacingset{1.5}

In Section A, we provide mathematical derivations for the identification formula. In Section B, we connect the VE parameters with a causal interpretation. In Section C, we show the details of VE estimation. In Section D, we provide a simulation study to demonstrate our statistical results.

\appendix

\section{Mathematical details for identification formula}
\subsection{Proof for Equation (1)}
The assumption $\Pr(R=\textup{Symptoms}, T=1|\mathcal{Z}) = H \times S \times f_s(\bX, V)$ implies that $I$ is independent of $T$ and $I\{R=\textup{Symptom}\}$ given $V, \bX, S=1, H=1$. Therefore,
\begin{align*}
    \Pr(I|V, \bX, S=1, R=\textup{Symptom}, H=1, T=1) &= \Pr(I|V, \bX, S=1, H=1) \\
    &=\frac{\Pr(I,S=1|V,\bX,H=1)}{\Pr(S=1|V,\bX,H=1)}.
\end{align*}
The odds ratio of observed data is then
\begin{align*}
    & \textup{Odds-ratio}_{I\sim V}(\bX, S=1, R=\textup{Symptom}, H=1, T=1) \\
    &=\frac{\Pr(I=1, S=1|V=1,\bX,H=1)}{\Pr( S=1|V=1,\bX,H=1)} \frac{\Pr(I=0, S=1|V=0,\bX,H=1)}{\Pr( S=1|V=0,\bX,H=1)} \\
    &\quad \times \frac{\Pr(S=1|V=0,\bX,H=1)}{\Pr(I=1, S=1|V=0,\bX,H=1)} \frac{\Pr(S=1|V=1,\bX,H=1)}{\Pr(I=0, S=1|V=1,\bX,H=1)} \\
    &= \frac{\Pr(I=1, S=1|V=1,\bX,H=1)\Pr(I=0, S=1|V=0,\bX,H=1)}{\Pr(I=1, S=1|V=0,\bX,H=1)\Pr(I=0, S=1|V=1,\bX,H=1)}.
\end{align*}
If we further assume that 
\begin{align*}
\Pr(I=0, S=1|V=0,\bX,H=1) =\Pr(I=0, S=1|V=1,\bX,H=1),
\end{align*}
then 
\begin{align*}
    \textup{Odds-ratio}_{I\sim V}(\bX,  S=1, R=\textup{Symptom}, H=1, T=1) = \frac{\Pr(I=1, S=1|V=1,\bX,H=1)}{\Pr(I=1, S=1|V=0,\bX,H=1)},
\end{align*}
which completes the proof.
\subsection{Proof for identification given case contact tracing}
The assumption $\Pr(R=\textup{Case contact tracing}, T =1|\mathcal{Z}) = C \times f_c(\bX, V)$ implies that $I$ is independent of $I\{R=\textup{Case contact tracing}\}, T$ given $V, \bX, C=1$. We then have
\begin{align*}
    &\Pr(I=1|V=1,\bX=\bx, C=1, R = \textrm{Case contact tracing}, T =1)\\
    &= \Pr(I=1|V=1,\bX=\bx, C=1),
\end{align*}
which implies the desired identification.

\subsection{Proof for connecting VE parameters}

We  first show when $VE_{symp}(\bx,1) = VE_{symp}(\bx)$. For $h \in \{0,1\}$, We have the following derivation:
\begin{align*}
    & \frac{\Pr(I=1,S=1|V=1,\bX, H = h)}{\Pr(I=1,S=1|V=0,\bX, H = h)} \Pr(H=h|V=0, \bX, I=1,S=1) \\
    &= \frac{\Pr(I=1,S=1,H=h|V=1,\bX)}{\Pr(I=1,S=1|V=0,\bX)} \frac{\Pr(H=h|V=0,X)}{\Pr(H=h|V=1,X)}  \\
 &=  \frac{\Pr(I=1,S=1,H=h|V=1,\bX)}{\Pr(I=1,S=1|V=0,\bX)}
\end{align*}
where the last step uses the conditional independence of $V$ and $H$ given $\bX$. Then, we have
\begin{align*}
    &\frac{\Pr(I=1,S=1|V=1,\bX)}{\Pr(I=1,S=1|V=0,\bX)} \\
    &=\frac{\Pr(I=1,S=1, H=1|V=1,\bX)}{\Pr(I=1,S=1|V=0,\bX)} + \frac{\Pr(I=1,S=1, H=0|V=1,\bX)}{\Pr(I=1,S=1|V=0,\bX)} \\   
    &= \frac{\Pr(I=1,S=1|V=1,\bX, H=1)}{\Pr(I=1,S=1|V=0,\bX, H=1)} \Pr(H=1|V=0, \bX, I=1,S=1) \\
    &\quad + \frac{\Pr(I=1,S=1|V=1,\bX, H=0)}{\Pr(I=1,S=1|V=0,\bX, H=0)} \Pr(H=0|V=0, \bX, I=1,S=1) .
\end{align*}
If we assume $H$ is not an effect modifier, i.e., 
\begin{align*}
\frac{\Pr(I=1,S=1|V=1,\bX, H=1)}{\Pr(I=1,S=1|V=0,\bX, H=1)} = \frac{\Pr(I=1,S=1|V=1,\bX, H=0)}{\Pr(I=1,S=1|V=0,\bX, H=0)},
\end{align*}
then we can see that
\begin{align*}
    \frac{\Pr(I=1,S=1|V=1,\bX)}{\Pr(I=1,S=1|V=0,\bX)} = \frac{\Pr(I=1,S=1|V=1,\bX, H=1)}{\Pr(I=1,S=1|V=0,\bX, H=1)},
\end{align*}
indicating $VE_{symp}(\bx,1) = VE_{symp}(\bx)$. For case contact tracing, the proof is similar by substituting $C$ for $H$.

\section{Causal interpretation of VE parameters}
We start from $VE_{symp}(\bx,1)$. For notation convenience, we define $D=I\{I=1,S=1\}$, with $D=1$ indicating having symptomatic infections.  We then define $D(v)$ as the potential outcome of a subject if their vaccination status were $V=v$ for $v=0,1$. For a causal interpretation of $VE_{symp}(\bx,1)$, we define, for $h = 0,1$,
 \begin{equation*}
     \textup{causal-}VE_{symp}(\bx,h) = 1-\dfrac{\Pr(D(1)=1\mid \bX = \bx, H=h)}{\Pr(D(0)=1\mid \bX = \bx, H=h)}.
 \end{equation*}
 {If $\textup{causal-}VE(\bx,h) = VE(\bx,h)$, then $VE(\bx,h)$ describes the direct effect of the vaccine against symptomatic infection by the virus of interest. This causal interpretation requires standard assumptions for causal inference~\citep{hernan2010causal}. First, we assume no interference, i.e., the potential outcomes of a subject are independent of the vaccination status of other subjects (reasonable in large populations). Second, we assume causal consistency, the observed outcome equals the potential outcome under the actual vaccination status, i.e. $D(v)=D$ if $V=v$ for $v=0,1$.
Third, we assume no other unmeasured confounding, i.e. $D(v)$ is independent of $V$ given $(\bX, H)$. This assumption requires that 
the only unmeasured confounder is healthcare-seeking behavior. 
Last, we assume ``positivity'': the population of interest has partial vaccine coverage within each confounder stratum, i.e. $0<\Pr(V=1\mid X=\bx, H=h)<1$ for every $\bx$ and $h$. 
}

To provide a causal interpretation for $VE_{symp}(\bx)$, we have two approaches. The first approach assumes that healthcare-seeking is not a confounder to obtain
\begin{align*}
   { VE_{symp}(\bx) = \textup{causal-}VE_{symp}(\bx) = 1 -\dfrac{\Pr(D(1)=1\mid \bX = \bx)}{\Pr(D(0)=1\mid \bX = \bx)}}
\end{align*}
under the same causal assumptions for $VE_{symp}(\bx, h)$.
A second approach assumes the same causal assumptions for $VE_{symp}(\bx,h)$ and uses proximal learning \citep{li2024double} to address the issue of unmeasured confounders by leveraging measured proxy variables such as vaccine history.

For $VE(\bx)$, we define $I(v)$ as the potential outcome of a subject if their vaccination status were $V=v$ for $v=0,1$, and let
\begin{align*}
   \textup{causal-}VE(\bx) = 1 -\dfrac{\Pr(I(1)=1\mid \bX = \bx)}{\Pr(I(0)=1\mid \bX = \bx)}.
\end{align*}
Then, achieving a causal interpretation of $VE(\bx)$, i.e., $VE(\bx)=\textup{causal-}VE(\bx)$ requires the same procedure and assumptions as for $VE_{symp}(\bx)$ above. 

Finally, for $VE_{cct}(\bx)$, we define $I(v,c)$ as the potential outcome given vaccination status $v$ and case-contact status $c$. This potential outcome is different from $I(v)$ because we are conditioning on the post-randomization variable $C=1$. Then, the  causal controlled direct effect is defined as
\begin{align*}
   \textup{CDE-}VE_{cct}(\bx) = 1 -\dfrac{\Pr(I(1,1)=1\mid \bX = \bx)}{\Pr(I(0,1)=1\mid \bX = \bx)}.
\end{align*}
In the epidemiology context, it refers to the VE unexplained by the behavior change after vaccination and can be complimentary to the primary VE.
To identify this causal quantity, we assume standard assumptions in causal mediation analysis \citep{vanderweele2011controlled}. First, we assume
no interference, i.e., the potential outcomes of a subject are independent of the vaccination status of other subjects (reasonable in large populations). Second, we assume causal consistency, $I(v,c) = I$ if $V = v$ and $C=c$. Third, we assume no unmeasured confounding: (1) $I(v,c)$ is independent of $C$ given $V,\bX$ and (2) $I(v,c)$ is independent of $V$ give $\bX$. Here, (1) is additionally required to include all confounders for the mediator-outcome relationship. If the healthcare-seeking behavior is considered as an unmeasured confounder, this assumption needs further conditioning on $H$ and using proximal learning approaches to address unmeasured confounding. Finally, we assume positivity for the distribution of $(C,V)$ given $\bX$.


\section{Details for estimation of $h(V,\bX,r)$ with asymptotic results}
In this section, we provide theoretical guarantees on the asymptotic distribution of our estimator based on different test reasons. We present the distribution of our estimator in both parametric and non-parametric settings in Theorem \ref{thm1} and Theorem \ref{thm2}.
\begin{assumption}\label{ass1}
We assume 
\begin{align*}
    \log(\Pr(I=1 |V,\bX,R=r,T=1))=h_{r}(V,\bX,\beta^*).
\end{align*}
This is equivalent to assuming $\Pr(I=1 |V,\bX,R=r,T=1)=\exp(h_{r}(V,\bX,\beta^*)).$
\end{assumption}

\begin{assumption}\label{ass2}
    We assume $h_r(1,\bX,\beta^*)-h_r(0,\bX,\beta^*)$ is independent with $r.$ This assumption is implied by the data-generating distribution for each of the three reasons for testing we consider.
\end{assumption}

\begin{assumption}\label{ass3}
    We assume we collect $n_r$ data for every sub-category and have $n$ observed data.
\end{assumption}
Estimation Procedure. For every dataset with label $R=r,$ we estimate $\beta^*$ using the technique for estimating generalized linear model (GLM) $\hat\beta_r.$

We first analyze the asymptotic behavior of $\hat\beta_r.$ In specific, we have
\begin{align*}
\sqrt{n_r}[h_r(1,\bX,\hat\beta_r)-h_r(0,\bX,\hat\beta_r)-(h_r(1,\bX,\beta^*)-h_r(0,\bX,\beta^*))]\rightarrow N(0,v_r^\top \Sigma_{0,r}v_r),
\end{align*}
where \begin{align*}
    \Sigma_{0,r}(\beta^*)=\textrm{Var}\bigg(\frac{\partial h_r}{\partial \beta}(V,\bX,\beta^*)(I-\exp(h_r(V_i,\bX_i,\beta^*))\Big{|}R=r,T=1\bigg)^{-1},
\end{align*}
and $v_r=\frac{\partial h_r}{\partial \beta}(1,\bX,\beta^*)-\frac{\partial h_r}{\partial \beta}(0,\bX,\beta^*),r\in[K].$

We then construct our test statistics via meta-learning as follows:
\begin{align*}
T_s=\sqrt{n}\sum_{r=1}^{K}\omega_r[\exp(h_r(1,\bX,\hat\beta_r)-h_r(0,\bX,\hat\beta_r))-\exp((h_r(1,\bX,\beta^*)-h_r(0,\bX,\beta^*)))]
\end{align*}

Next, we present the main theorem when the link function is parametric.

\begin{theorem}\label{thm1}
    Under Assumptions \ref{ass1}, \ref{ass2}, Assumption \ref{ass3} and Delta method, we obtain test statistics
    \begin{align*}
T_s&:=\sqrt{n}\sum_{r=1}^{K}w_r[\exp(h_r(1,\bX,\hat\beta_r)-h_r(0,\bX,\hat\beta_r))-\exp((h_r(1,\bX,\beta^*)-h_r(0,\bX,\beta^*)))]\\&\rightarrow N\bigg(0,\sum_{r=1}^{K}w_r^2\frac{\Pr(R\in \{1,\cdots,K\}|T=1)}{\Pr(R=r | T=1)}v_r^\top\Sigma_{0,r}(\beta^*) v_r\cdot\exp(h_r(1,\bX,\beta^*)-h_r(0,\bX,\beta^*))^2\bigg),
    \end{align*}
    where 
\begin{align*}
    \Sigma_{0,r}(\beta^*)=\textrm{Var}\bigg(\frac{\partial h_r}{\partial \beta}(V,\bX,\beta^*)(I_i-\exp(h_r(V_i,\bX_i,\beta^*))\Big{|}R=r,T=1\bigg)^{-1},
\end{align*}
and $v_r=\frac{\partial h_r}{\partial \beta}(1,\bX,\beta^*)-\frac{\partial h_r}{\partial \beta}(0,\bX,\beta^*),r\in[K].$
\end{theorem}

\begin{proof}
According to the definition of the test statistic $T_s,$ we obtain
    \begin{align*}
    T_s&=\sqrt{n}\sum_{r=1}^{K}w_r[\exp(h_r(1,\bX,\hat\beta_r)-h_r(0,\bX,\hat\beta_r))-\exp((h_r(1,\bX,\beta^*)-h_r(0,\bX,\beta^*)))]  \\&=\sqrt{n}\sum_{r=1}^{K}w_r\exp(h_r(1,\bX,\beta^*)-h_r(0,\bX,\beta^*))\cdot\bigg[\bigg(\frac{\partial h_r}{\partial \beta}(1,\bX,\beta^*)-\frac{\partial h_r}{\partial \beta}(0,\bX,\beta^*)\bigg)^\top(\hat\beta_r-\beta^*)\bigg]\\&\qquad+o(1).
\end{align*}
The second equality follows directly from Taylor's expansion. 

As $\hat\beta_r$ is estimated via GLM based on samples with label $R=r,r\in [K],$ we obtain
\begin{align*}
\hat\beta_r-\beta^*&=\frac{1}{n_r}\Sigma_{0,r}(\beta^*)\sum_{R_i=r}\bigg\{\frac{\partial h_r}{\partial \beta}(V,\bX,\beta^*)(I_i-\exp(h_r(V_i,\bX_i,\beta^*))\\&\qquad-\mathbb{E}\bigg[\frac{\partial h_r}{\partial \beta}(V,\bX,\beta^*)(I_i-\exp(h_r(V_i,\bX_i,\beta^*))\Big{|} R_i=r,T_i=1\bigg]\bigg\}+O\Big(\frac{1}{n_r}\Big),
\end{align*}
where 
\begin{align*}
    \Sigma_{0,r}(\beta^*)=\textrm{Var}\bigg(\frac{\partial h_r}{\partial \beta}(V,\bX,\beta^*)(I_i-\exp(h_r(V_i,\bX_i,\beta^*))\Big{|}R=r,T=1\bigg)^{-1}.
\end{align*}
After plugging in the expansion of $\hat \beta_r$ for $r\in[k]$ into the expression of $T_s$, we obtain
\begin{align*}
    T_s\rightarrow_d N\bigg(0,\sum_{r=1}^{K}w_r^2\frac{\Pr(R\in \{1,\cdots,K\}|T=1)}{\Pr(R=r | T=1)}v_r^\top\Sigma_{0,r}(\beta^*) v_r \cdot\exp(h_r(1,\bX,\beta^*)-h_r(0,\bX,\beta^*))^2\bigg),
\end{align*}
where $v_r=\frac{\partial h_r}{\partial \beta}(1,\bX,\beta^*)-\frac{\partial h_r}{\partial \beta}(0,\bX,\beta^*),r\in[K].$
\end{proof}

Next, we discuss the situation where the link function is in a non-parametric form. We also impose several assumptions. 

\begin{assumption}\label{ass_nonpara}
 We denote $\mathbb{E}[I_iV_i|X,T=1,R=r]=m_1^{r,T}(X),\mathbb{E}[(1-I_i)V_i|X,T=1,R=r]=m_2^{r,T}(X),\mathbb{E}[I_i(1-V_i)|X,T=1,R=r]=m_3^{r,T}(X),\mathbb{E}[(1-I_i)(1-V_i)|X,T=1,R=r]=m_4^{r,T}(X)$ are in $C^{s-1}$, $(s\ge 1)$. In addition, we also assume $f(X |T,R)$ (density of the covariates) belongs to $C^{s-1}$, $(s\ge 1)$. 
\end{assumption}

\begin{theorem}\label{thm2}
    Under Assumptions \ref{ass3} and \ref{ass_nonpara}, we obtain test statistics
    \begin{align*}
       T_s&=\sqrt{nh}\sum_{r=1}^Kw_r\Big( \log \widehat{RR}^{r,T}(x)-\log RR^{r,T}(x)\Big)\\&\rightarrow N\bigg(0,\sum_{r=1}^{K}w_r^2\frac{\Pr(R\in \{1,\cdots,K\}|T=1)}{\Pr(R=r | T=1)f(x|R=r,T=1)}c_r(x)^\top \Sigma_r(x) c_r(x)\int K^2(u) du RR^{r,T}(x)^2\bigg),
    \end{align*}
    if we choose bandwidth $h=o(n^{-1/(2s+d)})$ with $d$ being the dimension of the covariates $\bX$. Here
    \begin{align*}
     \widehat{RR}^{r,T}(\bx)&=\frac{\hat m_1^{r,T}(x)}{\hat m_2^{r,T}(x)}\cdot \frac{\hat m_4^{r,T}(x)}{m_3^{r,T}(x)}\\&=\frac{\sum_{i:R_i=r,T=1}K(\frac{X_i-x}{h})I_iV_i}{\sum_{i:R_i=r,T=1}K(\frac{X_i-x}{h})V_i}\cdot \frac{\sum_{i:R_i=r,T=1}K(\frac{X_i-x}{h})(1-V_i)}{\sum_{i:R_i=r,T=1}K(\frac{X_i-x}{h})I_i(1-V_i)},
    \end{align*}
    \begin{align*}
    c_r(x)&=\Big(1/m_1^{r,T}(x),-1/m_2^{r,T}(x),1/m_4^{r,T}(x),-1/m_3^{r,T}(x)\Big),\\
    \Sigma_r(x)&=Cov(v |X=x,R=r,T=1).
\end{align*}
with $v=(I_iV_i,V_i,(1-V_i),I_i(1-V_i))$ and $f(x|R=r,T=1)$ being the conditional density of $x$ given $R=r,T=1.$
\end{theorem}

\begin{proof}
We are interested in the odds ratio estimand:
\begin{align*}
    RR(\bx)=&\frac{\Pr(I=1|V=1,\bX=\bx,R=r,T=1)}{\Pr(I=1|V=0,\bX=\bx,R=r,T=1)}\\&
    = \frac{\mathbb{E}[I_iV_i|X,T=1,R=r]}{\mathbb{E}[V_i|X,T=1,R=r]}\cdot \frac{\mathbb{E}[(1-V_i)|X,T=1,R=r]}{\mathbb{E}[I_i(1-V_i)|X,T=1,R=r]}
\end{align*}

Therefore by Assumption \ref{ass_nonpara}, we obtain estimator:
\begin{align*}
    \widehat{RR}^{r,T}(\bx)&=\frac{\hat m_1^{r,T}(x)}{\hat m_2^{r,T}(x)}\cdot \frac{\hat m_4^{r,T}(x)}{m_3^{r,T}(x)}\\&=\frac{\sum_{i:R_i=r,T=1}K(\frac{X_i-x}{h})I_iV_i}{\sum_{i:R_i=r,T=1}K(\frac{X_i-x}{h})}\bigg{/}\bigg[\frac{\sum_{i:R_i=r,T=1}K(\frac{X_i-x}{h})V_i}{\sum_{i:R_i=r,T=1}K(\frac{X_i-x}{h})}\bigg]\\&\qquad\cdot \frac{\sum_{i:R_i=r,T=1}K(\frac{X_i-x}{h})(1-V_i)}{\sum_{i:R_i=r,T=1}K(\frac{X_i-x}{h})}\bigg{/}\bigg[\frac{\sum_{i:R_i=r,T=1}K(\frac{X_i-x}{h})I_i(1-V_i)}{{\sum_{i:R_i=r,T=1}K(\frac{X_i-x}{h})}}\bigg].
\end{align*}
where $K(\cdot)$ is an $s$-order kernel.

We next derive the asymptotic distribution of $\log OR(x).$ Since we choose bandwidth small enough $(h=o(n^{-1/(2m+d)}))$ such that the variance order $O(\frac{1}{n_rh^d})$ of the kernel estimator dominates the bias term $O(h^{2s})$. Therefore, for any $R=r, $ we obtain
\begin{align*}
&\sqrt{n_rh}(m_1^{r,T}(x)-m_1(x))\rightarrow N\bigg(0,\frac{Var(I\cdot V |X=x,R=r,T=1)}{f(x|R=r,T=1)}\int K^2(u) du\bigg)\\
&\sqrt{n_rh}(m_2^{r,T}(x)-m_2(x))\rightarrow N\bigg(0,\frac{Var(V |X=x,R=r,T=1)}{f(x|R=r,T=1)}\int K^2(u) du\bigg)\\
&\sqrt{n_rh}(m_3^{r,T}(x)-m_3(x))\rightarrow N\bigg(0,\frac{Var(1-V |X=x,R=r,T=1)}{f(x|R=r,T=1)}\int K^2(u) du\bigg)\\
&\sqrt{n_rh}(m_4^{r,T}(x)-m_4(x))\rightarrow N\bigg(0,\frac{Var(I\cdot (1-V) |X=x,R=r,T=1)}{f(x|R=r,T=1)}\int K^2(u) du\bigg)
\end{align*}
where $f(x|R=r,T=1)$ is the conditional density of $x$ given $R=r,T=1.$

By the Delta's method, we obtain
\begin{align*}
    &\sqrt{n_rh}(\log{\widehat{RR}^{r,T}}(x)-\log RR^{r,T}(x))\\&=\frac{\hat m_1^{r,T}(x)-m_1^{r,T}(x)}{m_1^{r,T}(x)}-\frac{ m_2^{r,T}(x)-\hat m_2^{r,T}(x)}{m_2^{r,T}(x)}+\frac{ m_4^{r,T}(x)-\hat m_4^{r,T}(x)}{m_4^{r,T}(x)}-\frac{ \hat m_3^{r,T}(x)- m_3^{r,T}(x)}{m_3^{r,T}(x)}+o_p(1)
\end{align*}
Therefore, we have
\begin{align*}
    \sqrt{n_rh}(\hat{\log OR^{r,T}}(x)-\log OR^{r,T}(x))\rightarrow N\bigg (0, c_r(x)^\top\Sigma_r(x)c_r(x)\int K^2(u) du/f(x|R=r,T=1)\bigg).
\end{align*}
where 
\begin{align*}
    c_r(x)&=\Big(1/m_1^{r,T}(x),-1/m_2^{r,T}(x),1/m_4^{r,T}(x),-1/m_3^{r,T}(x)\Big),\\
    \Sigma_r&=Cov(v |X=x,R=r,T=1).
\end{align*}
where $v=(I_iV_i,V_i,(1-V_i),I_i(1-V_i)).$

Recall the test statistics is
\begin{align*}
    T_s=&\sqrt{nh}\sum_{r=1}^K w_r\Big(\exp(\log \widehat{RR}^{r,T}(x))-\exp(\log RR^{r,T}(x))\Big)\\&\rightarrow N\bigg(0,\frac{\sum_{r=1}^{K}w_r^2\frac{\Pr(R\in [K]|T=1)}{\Pr(R=r|T=1)}c_r(x)^\top \Sigma_r(x) c_r(x)\int K^2(u) du}{f(x|R=r,T=1)}\cdot RR^{r,T}(x)^2\bigg).
\end{align*}
\end{proof}

\section{Simulations}
Using simulations, we aim to demonstrate that the traditional odds ratio estimator can be biased for TNDs with multiple reasons for testing, with the proposed stratified analysis being unbiased for the target estimands (with the necessary assumptions), and quantifying the precision gained by leveraging data for asymptomatic reasons for testing.

\subsection{Simulation design}
We consider two observed covariates: age ($X_1$, scaled to be in $[0.5,1]$), and sex ($X_2$, as a binary variable), together with the latent healthcare-seeking behavior $H$.
We assume $X_1 \sim \textup{Uniform}([0.5,1])$, $X_2|X_1 \sim \textup{Bernoulli}(0.5)$, $H|(X_1,X_2) \sim \textup{Bernoulli}(X_1)$, $V|(X_1,X_2,H) \sim \textup{Bernoulli}(0.6X_1+0.1X_2)$ with an additional variable $C|(X_1,X_2,H,V) \sim \textup{Bernoulli}(0.05(1+X_2))$ representing an indicator of case contact. We then generate $I$ from a log-binomial model with
\begin{align*}
   \log\ \Pr(I=1|V,X_1,X_2,H,C) = \beta_0 - 0.5V + 0.5X_1 -X_2+0.1C,
\end{align*}
where $\beta_0$ is the intercept reflecting underlying infection prevalence.  
We also consider the full combination of the following scenarios: (a) age and healthcare-seeking behavior modify VE (replacing $-0.5V$ above by $-0.5VH-0.1VX_1$) or not; (b) the disease of interest is relatively rare ($\Pr(I=1) = 0.10$ setting $\beta_0 =-2.1$) or common ($\Pr(I=1) = 0.50$ setting $\beta_0 =-0.6$). 
Next, we generate $R$ and $T$ following the DAGs in Sections 2.1 and 2.3, with probability $0.2, 0.24, 0.07, 0.03$ of taking a test due to symptoms, mandatory screening, case contact tracing, and others respectively; and the probability of not taking a test is $0.46$.
Details for the distributions of $R$ and $T$ Specifically, given the generated $\mathcal{Z} = \{X_1, X_2, H, V, C, I\}$ and $I$, we generate $R$ and $T$ as follows:
\begin{align*}
    \Pr(R = \textup{Symptoms}, T=1 | \mathcal{Z}) &= 0.4(1-C)HX_1 G,\\
    \Pr(R = \textup{Mandatory screening}, T=1 | \mathcal{Z}) &= (0.2V+0.2 X_1)(1-C), \\
    \Pr(R = \textup{Case contact tracing}, T=1 | \mathcal{Z}) &=C(0.9+0.1 X_2),\\
    \Pr(R = \textup{Other}, T=1 | \mathcal{Z}) &=0.4(1-C)I(1-V),
\end{align*}
where 
\begin{align*}
    G &= (1-I)\left\{1-V + V \frac{\Pr(I=0|V=0,X_1,X_2,H,C=0)}{\Pr(I=0|V=1,X_1,X_2,H,C=0)} \right\} \\
    &\quad + I \cdot \Pr(I=0|V=0,X_1,X_2,H,C=0)
\end{align*}
such that $G$ makes the Equation (1) of the main paper hold and, meanwhile, makes the logistic regression model correctly specified.

For each scenario, we simulated 1000 data sets, each containing $10,000$ individuals reflecting  5400 tests on average. 
For each simulated data set, we compute the odds ratio estimator based on $R=\textup{Symptoms}$ (``Odds-Ratio-s'', serving as the benchmark) and
the odds ratio estimator naively aggregates all reasons for testing (``Odds-Ratio-all'') using the same logistic regression model. 
We also compute the stratified estimator based on mandatory screening and case contact tracing (``Stratified''), and the stratified estimator based on all three reasons (''Stratified-all'') with a correctly specified log-binomial model and inverse-variance weighting. 
Under the scenario without effect modification from covariates, the metrics of performance include bias, empirical standard error based on Monte Carlo replications, averages of estimated standard errors, and coverage probability of the true VE based on a Normal-approximated 95\% confidence interval. If VE varies by $X_1$ and $H$, we visualize the coverage probability of ``Odds-Ratio-s'' and ``Stratified'' according to their target  VE values.


\subsection{Results}\label{sec:result}
Under the scenarios with homogeneous VE across strata, simulation results are summarized in Table~\ref{tab1}, where $VE(\bx)$ is set equal to $VE(\bx,1) = 1-e^{-0.5}$. 
In Table~\ref{tab1}, the odds-ratio-s, stratified, and stratified-all estimators all have negligible bias and nominal coverage probability, whereas the odds-ratio-all estimator has large bias and $0\%$ coverage probability across scenarios. This result confirms that the odds ratio estimator mixing all reasons for testing can cause bias and severe under-coverage. In contrast, our proposed stratified estimators yield valid inferences on VE, as expected. In terms of precision, among unbiased estimators,  our stratified estimator combining all three reasons for testing has the smallest variance, which is as little as approximately 25\% of the variance of the classical odds ratio estimator based on symptomatic subjects only, demonstrating the substantial benefits of improved precision through including the additional data using a stratified analysis.  
Comparing the settings of high prevalence versus low prevalence of infection, the results are similar, except that fewer cases proportionally increase the variance of all estimators. 

   \begin{table}[ht]
	\begin{center}
		\def\arraystretch{1.1}
		\setlength\tabcolsep{8pt}
		\begin{tabular}{crrrrr}
  \hline
	\makecell[c]{Prevalance\\ of infection }	&Estimators  & Bias & \makecell[r]{Empirical\\ standard \\ error}  & \makecell[r]{Averaged\\ standard error \\ estimates} & \makecell[r]{Coverage\\ probability}   \\
			\hline
  \multirow{4}{*}{\makecell[c]{High \\ (50\%)}}  & Odds-ratio-s &-0.00 & 0.09 & 0.09 & 0.94 \\
    & Odds-ratio-all &0.40 & 0.01 & 0.01 & 0.00 \\  
     &Stratified   &0.00 & 0.02 & 0.02 & 0.94 \\ 
     &Stratified-all &0.00 & 0.02 & 0.02 & 0.94 \\  
    \hline
  \multirow{4}{*}{\makecell[c]{Low \\ (10\%)}}  & Odds-ratio-s &-0.01 & 0.10 & 0.10 & 0.95 \\ 
			& Odds-ratio-all&0.30 & 0.03 & 0.03 & 0.00 \\ 
			&Stratified &  0.01 & 0.07 & 0.07 & 0.95 \\ 
			&Stratified-all &  0.01 & 0.06 & 0.06 & 0.94 \\ 
\hline
		\end{tabular}
	\end{center}
	\caption{Results of estimated vaccine effect using different estimators under homogeneous VE across strata. 
    Odds-ratio-s: odds-ratio estimator based on symptoms. Odds-ratio-all: odds-ratio estimator pooling all reasons for testing. Stratified: stratified analysis combining mandatory screening and case contact tracing. Stratified-all: stratified analysis combining all three reasons for testing.}
	\label{tab1}
\end{table} 

\begin{figure}
    \centering
    \begin{subfigure}[b]{0.45\textwidth}
        \includegraphics[width=\textwidth]{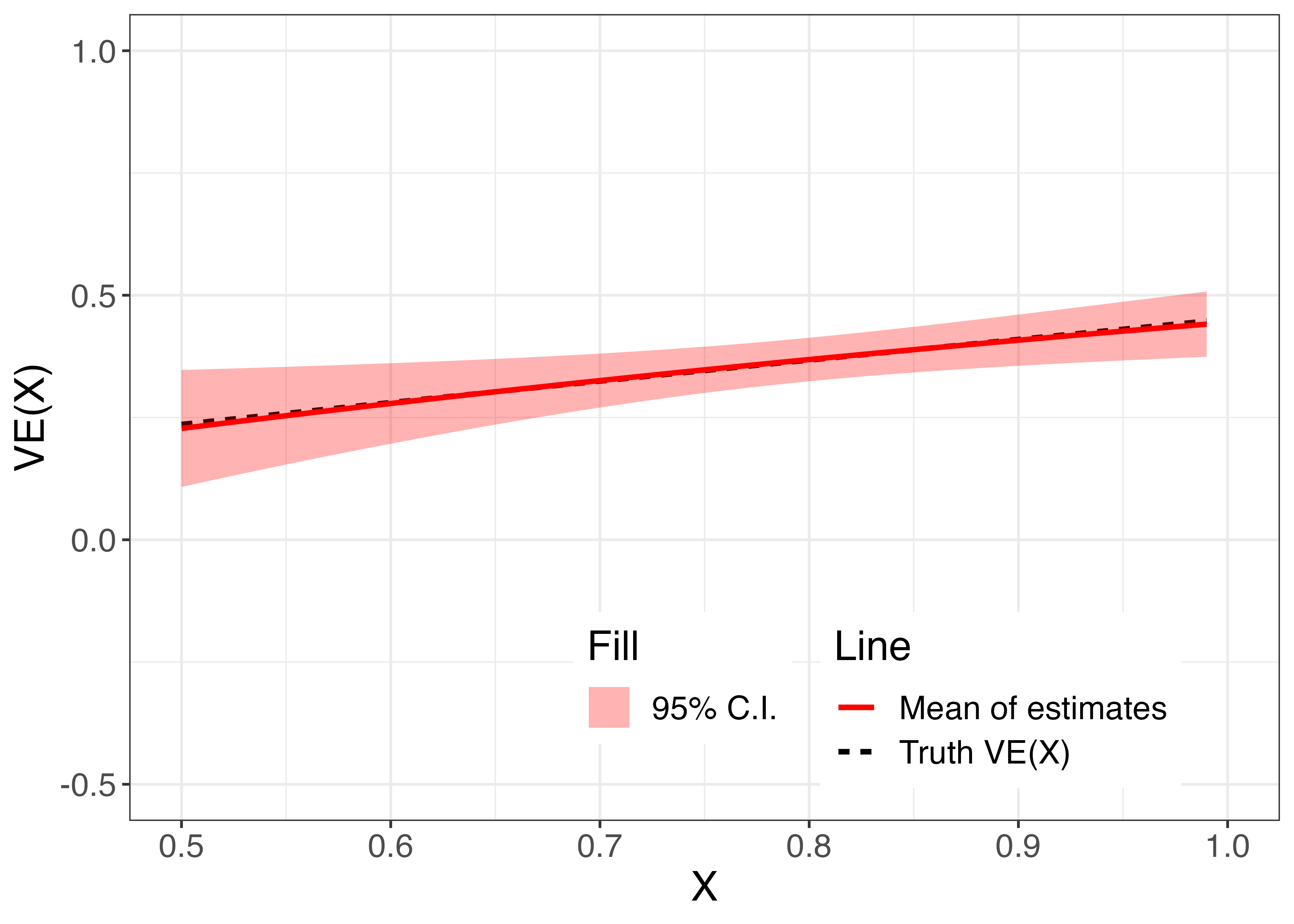}
        \caption{Estimating $VE(X_1)$ under high\\ infection prevalence.}
    \end{subfigure}
    \begin{subfigure}[b]{0.45\textwidth}
        \includegraphics[width=\textwidth]{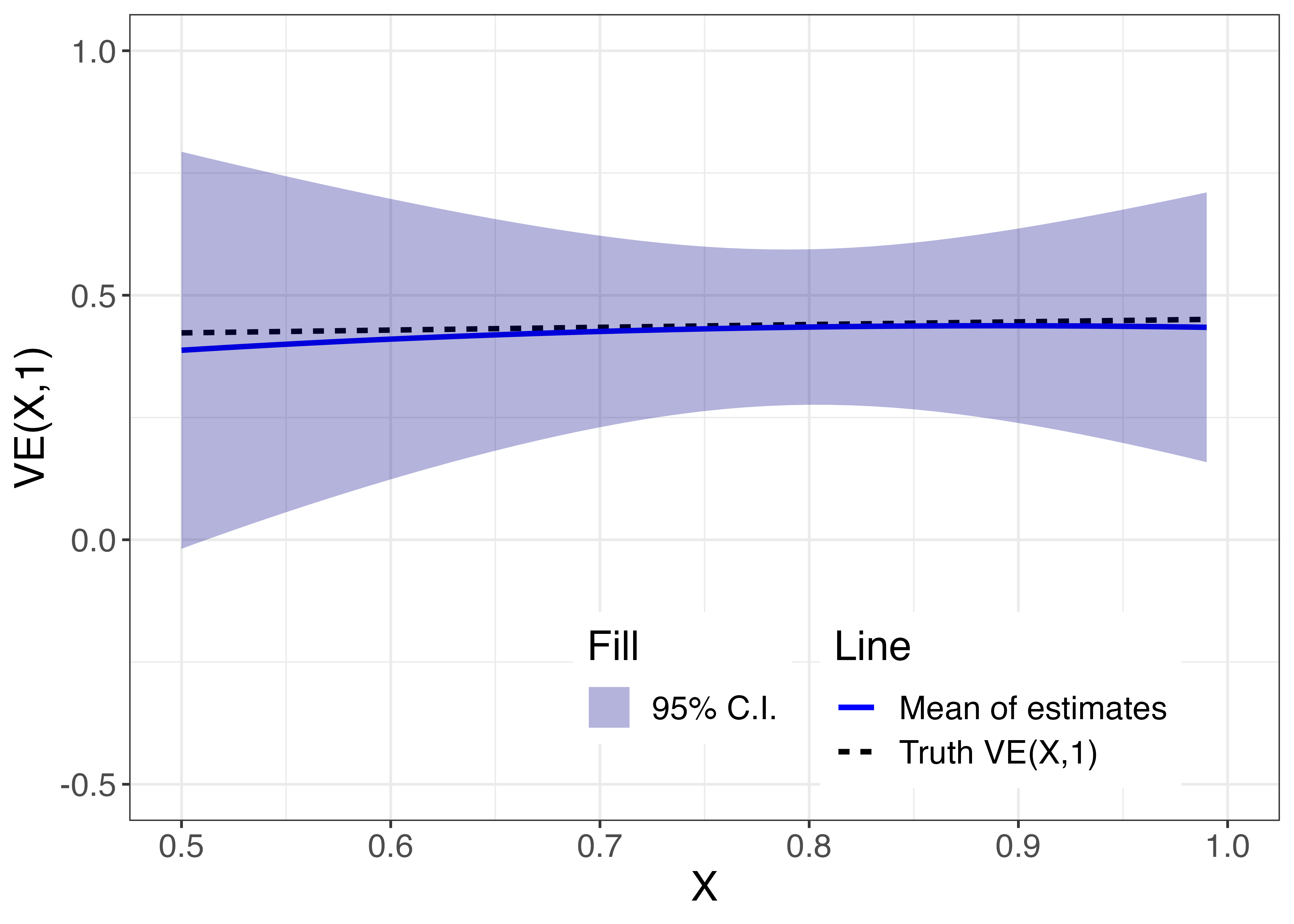}
        \caption{Estimating $VE(X_1,1)$ under high\\ infection prevalence.}
    \end{subfigure}
    \begin{subfigure}[b]{0.45\textwidth}
        \includegraphics[width=\textwidth]{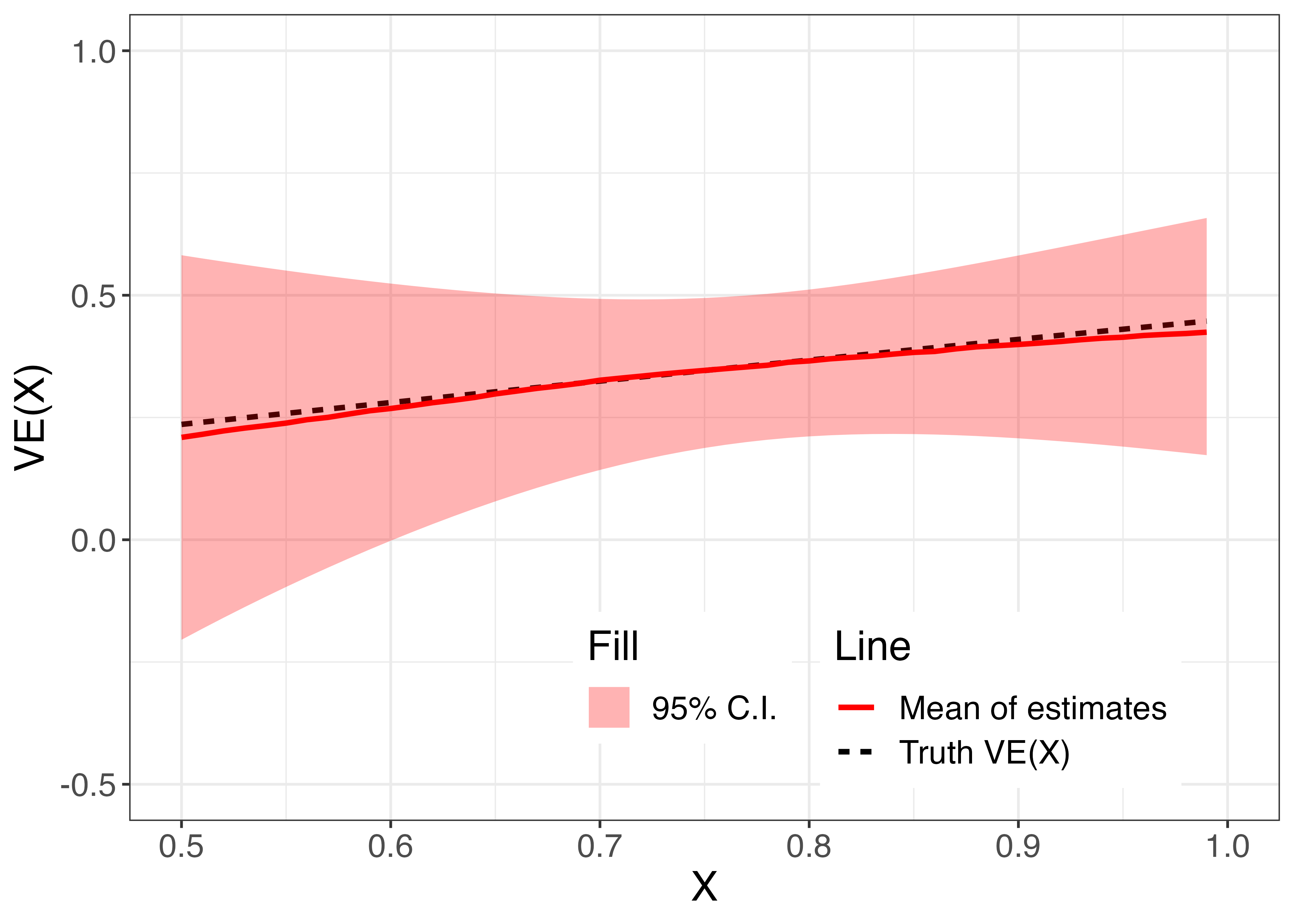}
        \caption{Estimating $VE(X_1)$ under low\\ infection prevalence.}
    \end{subfigure}
    \begin{subfigure}[b]{0.45\textwidth}
        \includegraphics[width=\textwidth]{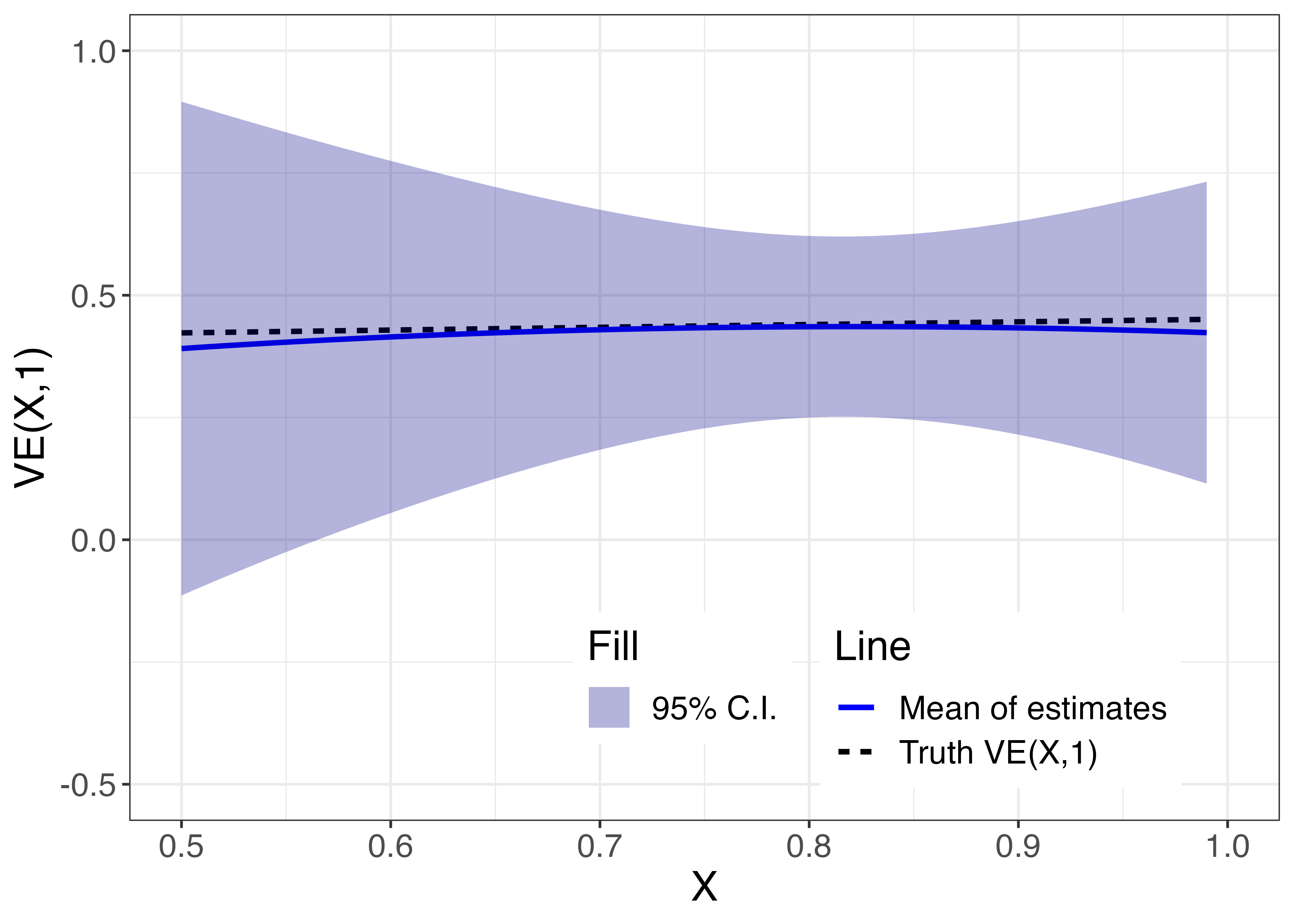}
        \caption{Estimating $VE(X_1,1)$ under low\\ infection prevalence.}
    \end{subfigure}
    \caption{Results of heterogeneous VE estimation as a function of $X_1$. The dashed black lines are the true VEs, which largely overlap with the colored solid lines, which are the mean of estimates. The colored areas are averaged 95\% confidence intervals across simulation replicates. 
    The left panels (a and c) reflect the stratified estimators based on mandatory screening and case contact tracing, while the right panels reflect the odds ratio estimator based on symptoms. }\label{fig:sim1}
\end{figure}

Figure~\ref{fig:sim1} presents scenarios where the VE varies by age and healthcare-seeking behavior with $VE(X_1, 1) = e^{-0.5-0.1X_1}$ and $VE(X_1) = e^{-0.5-0.1X_1}X_1 + e^{-0.1X_1}(1-X_1)$, both are increasing functions of $X_1$. Since $VE(X_1, 1) \ge VE(X_1)$, we use the odds-ratio-s estimator to estimate $VE(X_1, 1)$ and the stratified estimator to estimate $VE(X_1)$; however,  neither odds-ratio-all or stratified-all estimator is valid for this purpose. Figure~\ref{fig:sim1} shows that our conditional VE estimation is able to capture the truth across scenarios, reflected by the high overlap between the mean of estimates and the true values. This result demonstrates the potential of our proposed method to estimate heterogeneous VE and construct appropriate 95\% confidence bands. 


In addition to these two simulations, we have performed supplementary simulations to explore the effect of misclassification of the infection status and missing values in the reasons for testing, which are provided in Web Appendix E. Under our simulation settings, we found that 10\% misclassification of the infection status can cause negative bias ranging from $-0.06$ to $-0.21$. Missing values in the reasons for testing will not cause bias under the missing completely at random assumption, while 20\% missingness can inflate the standard error by around {12\%}.


\subsection{Additional Simulation Results}\label{add_simu}

In this section, we study three additional simulation results.  We first consider the misclassification of testing outcomes $I$. Specifically, we assume $10\%$ percentage of $D$ is misclassified. The results are presented in Table \ref{tab: misclassification rate}.
Second, we consider the situation where some of the testing reasons are missing or not specified. We let $20\%$ of each category be missing. The results are presented in Table \ref{tab: missing reasons}. 

   \begin{table}[ht]
	\begin{center}
		\def\arraystretch{1.1}
		\setlength\tabcolsep{8pt}
		\begin{tabular}{crrrrr}
  \hline
	\makecell[c]{Prevalance\\ of infection }	&Estimators  & Bias & \makecell[r]{Empirical\\ standard \\ error}  & \makecell[r]{Averaged\\ standard error \\ estimates} & \makecell[r]{Coverage\\ probability}   \\
			\hline
  \multirow{4}{*}{\makecell[c]{High \\ (50\%)}}  & Odds-ratio-s &-0.19 & 0.11 & 0.11 & 0.61 \\
    & Odds-ratio-all & 0.33 & 0.02 & 0.02 & 0.00  \\  
     &Stratified   &-0.06 & 0.03 & 0.03 & 0.28  \\ 
     &Stratified-all &-0.07 & 0.02 & 0.02 & 0.18 \\  
    \hline
  \multirow{4}{*}{\makecell[c]{Low \\ (10\%)}}  & Odds-ratio-s &-0.21 & 0.10 & 0.10 & 0.43\\
			& Odds-ratio-all&0.19 & 0.03 & 0.03 & 0.00 \\ 
			&Stratified &-0.19 & 0.07 & 0.07 & 0.12 \\  
			&Stratified-all &  -0.19 & 0.05 & 0.06 & 0.04 \\ 
\hline
		\end{tabular}
	\end{center}
	\caption{Results of estimated vaccine effect assuming 10\% misclassification rate and homogeneous VE across strata. 
    Odds-ratio-s: odds-ratio estimator based on symptoms. Odds-ratio-all: odds-ratio estimator pooling all reasons for testing. Stratified: stratified analysis combining mandatory screening and case tracing. Stratified-all: stratified analysis combining all three reasons for testing.}
	\label{tab: misclassification rate}
\end{table}     
   \begin{table}[ht]
	\begin{center}
		\def\arraystretch{1.1}
		\setlength\tabcolsep{8pt}
		\begin{tabular}{crrrrr}
  \hline
	\makecell[c]{Prevalance\\ of infection }	&Estimators  & Bias & \makecell[r]{Empirical\\ standard \\ error}  & \makecell[r]{Averaged\\ standard error \\ estimates} & \makecell[r]{Coverage\\ probability}   \\
			\hline
  \multirow{4}{*}{\makecell[c]{High \\ (50\%)}}  & Odds-ratio-s &0.00 & 0.10 & 0.10 & 0.94 \\
    & Odds-ratio-all & 0.40 & 0.02 & 0.02 & 0.00  \\  
     &Stratified   &-0.00 & 0.03 & 0.02 & 0.95  \\ 
     &Stratified-all &0.00 & 0.02 & 0.02 & 0.95 \\  
    \hline
  \multirow{4}{*}{\makecell[c]{Low \\ (10\%)}}  & Odds-ratio-s &-0.01 & 0.11 & 0.11 & 0.95\\
			& Odds-ratio-all&0.30 & 0.03 & 0.03 & 0.00\\ 
			&Stratified &0.01 & 0.08 & 0.08 & 0.93 \\  
			&Stratified-all &  0.02 & 0.07 & 0.06 & 0.92 \\ 
\hline
		\end{tabular}
	\end{center}
	\caption{Results of estimated vaccine effect assuming 20\% missing reasons for testing and homogeneous VE across strata. 
    Odds-ratio-s: odds-ratio estimator based on symptoms. Odds-ratio-all: odds-ratio estimator pooling all reasons for testing. Stratified: stratified analysis combining mandatory screening and case tracing. Stratified-all: stratified analysis combining all three reasons for testing.}
	\label{tab: missing reasons}
\end{table} 



For the first two scenarios outlined above, our focus remains similarly on the study of common and rare diseases under conditions where health-seeking behavior does not introduce bias to the vaccine effect, since the conclusions drawn under these conditions can be easily extrapolated to situations with biased vaccine effects due to health-seeking behavior.

Based on the results presented, we find that in the presence of a misspecified \(D\), the estimated odds ratio exhibits bias. Both the estimated mean and variances deviate from their theoretical values. A higher proportion of misclassification results in greater biases.  When there exist missed testing reasons, as the missing ratio for each reason increases, the standard error of the estimators expands in comparison to cases with no missing data.

{
\bibliographystyle{apalike}
\bibliography{references}
}